\documentclass[%
 aip,
 floatfix,
 amsmath,amssymb,
 reprint,%
]{revtex4-1}

\usepackage{graphicx}
\usepackage{dcolumn}
\usepackage{bm}

\usepackage[utf8]{inputenc}
\usepackage[T1]{fontenc}
\usepackage[english]{babel}
\usepackage{mathptmx}
\usepackage{etoolbox}
\usepackage{xcolor}
\usepackage{soul}

\makeatletter
\def\@email#1#2{%
 \endgroup
 \patchcmd{\titleblock@produce}
  {\frontmatter@RRAPformat}
  {\frontmatter@RRAPformat{\produce@RRAP{*#1\href{mailto:#2}{#2}}}\frontmatter@RRAPformat}
  {}{}
}%
\makeatother
\begin{document}

\preprint{AIP/123-QED}

\title[Low-dimensional representation of intermittent geophysical turbulence with H-SiNN]{Low-dimensional representation of intermittent geophysical turbulence with High-Order Statistics-informed Neural Networks (H-SiNN)}
\author{R. Foldes}
 \email{raffaello.foldes@ec-lyon.fr}
\affiliation{Universit\'e de Lyon, CNRS, Ecole Centrale de Lyon, INSA de Lyon, Universit\'e Claude Bernard Lyon 1, Laboratoire de M\'ecanique des Fluides et d’Acoustique - UMR 5509, F-69134 Ecully, France}
\affiliation{ 
Dipartimento di Scienze Fisiche e Chimiche, Universit\`a dell'Aquila, 67100 Coppito (AQ), Italy
}
\author{E. Camporeale}%
\affiliation{ 
Department of Physics and Astronomy, Queen Mary University of London, London E1 4NS, United Kingdom
}
\affiliation{
CIRES, University of Colorado, Boulder, CO, USA
}
\affiliation{
NOAA Space Weather Prediction Center, Boulder, CO, USA
}
\author{R. Marino}
 \affiliation{Universit\'e de Lyon, CNRS, Ecole Centrale de Lyon, INSA de Lyon, Universit\'e Claude Bernard Lyon 1, Laboratoire de M\'ecanique des Fluides et d’Acoustique - UMR 5509, F-69134 Ecully, France}

\date{\today}

\begin{abstract}
We present a novel machine learning approach to reduce the dimensionality of state variables in stratified turbulent flows governed by the Navier-Stokes equations in the Boussinesq approximation. The aim of the new method is to perform an accurate reconstruction of the temperature and the three-dimensional velocity of geophysical turbulent flows developing non-homogeneities, starting from a low-dimensional representation in latent space, yet conserving important information about non-Gaussian structures captured by high-order moments of distributions. To achieve this goal we modify the standard Convolutional Autoencoder (CAE) by implementing a customized loss function that enforces the accuracy of the reconstructed high-order statistical moments. We present results for compression coefficients up to 16 demonstrating how the proposed method is more efficient than a standard CAE in performing dimensionality reduction of simulations of stratified geophysical flows characterized by intermittent phenomena, as observed in the atmosphere and the oceans.
\end{abstract}

\maketitle


\section{\label{sec:level1}Introduction}

Geophysical fluids are characterized by the interplay of non-linear vortices and waves, developing very complex turbulent dynamics due to the density stratification and the Earth's solid body rotation.
Large-scale motions at planetary scales (of the order of 1000 km) emerge from the so-called quasi-geostrophic balance~\citep{Vallis2017}. At mesoscale $O(100\,\,\rm km)$ and even more at the sub-mesoscale $O(10\,\,\rm km)$, when the turbulent eddy turnover time becomes comparable to the characteristic time scales associated to internal waves, motions feel less the constrain of the force balance; a regime generically identified as \textit{stratified turbulence}~\citep{Lilly1983}. This is characterized by the presence of shear {layers} leading to instabilities, the vertical extension of strata being controlled by the Brunt-V\"ais\"al\"a frequency $N$. In stratified geophysical flows{,} the probability density function (PDF) of {the fields} at the scales comparable {to} that of the mean flow can be characterized by fat tails with a departure from Gaussianity; {in literature we refer to this phenomenology as \emph{large-scale intermittency}~\citep{Feraco2018,Feraco2021,Marino2022}, which differ from the classical (internal) small-scale intermittency detected as a departure from Gaussianity of the statistics of the field gradients. Large-scale non-Gaussian distributions of the dynamical fields are observed also in other fluid frameworks in nature, such as in the solar wind, where local shears associated to large values of large-scale increments of velocity and density are thought to trigger magnetohydrodynamic turbulent cascades in space plasmas \cite{marino2012,marino2023}.} {Such peculiar} behavior {of stratified turbulent flows} has been observed in the vertical velocity and (potential) temperature in both atmosphere \citep{Mahrt1987,Mahrt1988,Lenschow2011,Chau2021} and oceans \citep{DAsaro2007,Capet2008}, and {was} extensively {characterized} in a variety of numerical investigations \cite{Rorai2014,Feraco2018,Sujovolsky2019,Feraco2021,MDPI2021,Marino2022} through the forth-order moment of the vertical velocity (i.e., {through its} kurtosis). \citet{Marino2022} have shown how the extreme drafts, responsible for the non-Gaussian behavior of the vertical component of the velocity, do generate local turbulence and enhance the {internal} small-scale intermittency~\citep{Feraco2021}. They {provided as well} evidence that these structures are associated with patches of enhanced kinetic and potential energy dissipation, making stably stratified flows {in a certain parameter space} (analyzed in~\citet{Feraco2018}) more efficient at dissipating energy {and less homogeneous} due to the {irregular occurrence (in space and time)} of {these large-scale intermittent drafts}.
The community resorted to a variety of different techniques to model {the wide range of scales over which dynamics develop} in geophysical flows, from the implementation of codes based on Reynolds averaged Navier-Stokes (RANS) equations and Large Eddy Simulations schemes (LES)~\citep{Spalart1992,Girimaji2005,Spalart2009}, to the simulation of {rotating flows} using {reduced models}~\citep{Julien1996}. 
On the other hand, {progress} made in the field of {High-Performance} Computing {has} made it possible to perform direct numerical simulations (DNS) of stratified turbulent flows in a parameter space of geophysical interest \citep{pouquet2013,marino2015,rosenberg2015,oks}. 
{Regardless of } the physical model implemented, {a} common trait of high-resolution three-dimensional simulations of geophysical flows is the large amount of data produced. For example, a DNS run with a resolution of $512^3$ grid points in single precision requires more than 0.5{GB} of storage space per field and per time-step. Retaining the {simulation} output {with high temporal cadence} is therefore in many {cases} not doable, which requires {to find} a {trade off} though investigating the evolution of the fields in space and time is often needed to assess the complex dynamics developing in geophysical flows. The implementation of low-dimensional representations of the physical fields {appears on the one hand to be} a viable option for the creation of database allowing for the post-processing of {massive} data, relieving from the need to perform run-time analysis, {ashearsnd on the other is useful} for the {usage} of machine learning techniques able to {assimilate} features of the dynamical systems {directly} from the latent space~\citep{PARENTE2009,Wan2018,DeJesus2023}. 
The rationale behind the dimensionality reduction of the output of turbulent {flow} simulations {serves therefore multiple purposes, storing} data for post-processing at higher temporal frequency, {enabling} the efficient training of data-mining and machine learning techniques; {implementing forecasting tools able to operate using coarse-grained descriptions of the system~\citep{Champion2019,Bukka2021}.} The study of {low-dimensional} manifolds has a long history in {the context of} dynamical systems and turbulence
~\citep{Sirovich1987,Zdybal2022}, laying at the foundation of so-called reduced order methods~\citep{Lassila2014}. {The latter} have also recently benefited {from progresses made} in machine learning based approaches \citep{mohan2018deep,chen2021physics}.
Dimensionality reduction techniques such as proper orthogonal decomposition (POD) \citep{Berkooz1993, Holmes1996,Hamidreza2022,Morgan2023} and dynamic mode decomposition \citep{Schmid2010, Kutz2016} have been extensively used to address issues concerning a variety of flows~\citep{Berkooz2003,Garicano2019,Williams2013,marensi2023,Cheng2021}. 
Other examples of model order reduction techniques applied to fluid frameworks are Galerkin-projection based nonlinear methods \citep{Carlberg2013,MAULIK2020} and system identification based auto-regressive models \citep{Raveh2004, Sarma2017}. {Techniques based on machine learning principles that have been employed in modeling turbulent flows, in very different contexts, are Gaussian process regression \citep{Raissi2016, Zhang2015}, symbolic regression \citep{Brunton2016, Schmelzer2020}, field inversion \citep{Duraisamy2015, Parish2016}, artificial neural networks (ANNs)~\citep{Kochkov2021} and many others.} In data-driven methods, ANNs are the most powerful tools in terms of their ability to generalize and to capture highly nonlinear phenomena, typical of turbulent processes. 
{Other developments have focused on improving the reliability and accuracy of low fidelity models (i.e., RANS and LES), by using data from high fidelity simulations, e.g. DNS, either to learn proper closures~\citep{Charalampopoulos2022,McConkey2022,Srinivasan2023} or to generate high-resolution synthetic turbulent states starting from a coarser description of the flow~\citep{Angriman2023}}. Convolutional neural networks (CNNs) have been successfully applied to the identification of flow structures~\citep{Morimoto2021} and to perform nonlinear modal decomposition in turbulent flows~\citep{Fukami2020,murata2020}; more recently, CNN-based architectures have been extended to fully 3D DNS of turbulent channel flow~\citep{Nakamura2021} and numerical simulations of flows characterized by the emergence of extreme events, in terms of boundary coherent structures~\citep{Khoa2023}. These implementations emphasized the advantages of using CNNs over traditional methods based on principal component analysis (PCA), the former being able to capture the intrinsic nonlinear dynamics of turbulent flows. More generally, deep learning proves to be a powerful tool for the analysis and generation of reduced-order models of turbulent systems. In terms of machine learning applications to stratified flows, CNN-based deep learning has recently been applied by~\citet{Salehipour2019} to {obtain a} parametrization of diapycnal mixing using data from DNS. 
Convolutional AutoEncoders (CAE) are a family of autoencoders that proved successful in extracting information from two- and three-dimensional data \citep{Wang2016} and their use has recently been proposed in the context of fluid-dynamics \citep{Gonzalez2018, Xu2019, King2018}. However, most of these approaches present severe limitations if implemented without proper tuning, and their use is often constrained to idealized and/or simplified cases due to the major constraints that come from capturing transient behavior~\citep{Kutz2017}. Conversely, the presence of large-scale transient phenomena developing in geophysical flows, such as hurricanes, tornadoes, oceanic fronts, etc., can be viewed as an obstacle to the creation of reduced-order manifolds sufficiently informative to reliably recover properties and dynamical features of geophysical fluid systems. Multi-scale models, such as global circulation models (GCM), are powerful tools for {investigating} the Earth's atmosphere and the oceans which would greatly benefit from the possibility {of simulating} complex fluid dynamics on a dimensional phase space lower than what can be achieved numerically. 
The efficiency of machine learning approaches is reduced when applied to systems with strong transients, such as the non-stationary states resulting from the presence of intermittent phenomena in turbulent flows. The purpose of the study presented here is precisely to develop a machine learning tool based on CAE able to produce reliable low-order representations of stratified geophysical turbulent flows in a range of parameters characterized by {the development of} intermittent (in space and time) vertical velocity drafts, leading to the departure of the vertical velocity statistical moments from the Gaussian reference. 
We show that standard CAE architectures are well suited to learn a latent space containing enough information to reconstruct the general features of the original stratified turbulent flow. However, strong vertical drafts developing in a certain regime of the governing parameter, and in general {high-order} statistics of velocity and temperature in {the} presence of large- and small-scale intermittency, will typically not be well recovered.
To overcome this issue, we introduce here a novel implementation of CAE: statistics-informed convolutional autoencoder (or SiCAE). The idea is based on a more general approach, applicable to any neural network, which we refer to as high-order statistics-informed neural network (H-SiNN). The concept of statistics-informed neural network (SiNN) {was} first introduced by~\citet{ZHU2023}, where it has been proposed the addition of two terms to the loss function whose effect is to constrain both the PDF and the auto-correlation function of the reconstructed fields. On the other hand, in our implementation, the loss function explicitly enforces the preservation of high-order moments and this is done in order to ensure consistency between the statistics of the original and the reconstructed flow fields. Indeed, it is worth recalling that statistical moments of the velocity distribution function play a crucial role in the characterization of turbulent frameworks, being directly related to invariants such as energy and enstrophy, {and also} to other fundamental quantities, such as the dissipation occurring at the small scales.
{We will demonstrate here the enhanced capability of the H-SiNN, with respect to a standard CAE implementation, in recovering finer statistical features as well as flow inhomogeneities and intermittent dynamics.}\\
The manuscript is organized as follows: in Sec.~\ref{sec:sec1} the numerical simulations performed and analyzed are briefly introduced along with a description of the equations and parameters governing the dynamics of the stratified flows under study; Sec.~\ref{sec:cae_basics} explains the classical CAE architecture, and in Sec.~\ref{sec:sec4} our implementation of statistical-informed CAE is thoroughly described; finally, Sec.~\ref{sec:results} shows in detail the outcome of the application of the model reduction tool we developed to DNS of stratified turbulent flows developing large-scale intermittent events in the vertical component of the velocity and in the temperature field \citep{Feraco2018}.

\section{\label{sec:sec1}DNS solver for Stratified Turbulence}

The CAEs are trained using the output of a direct numerical simulation of a stably stratified turbulent flow obtained by integrating the Navier-Stokes equations in the Boussinesq approximation in which the velocity field ${\bf u}=({\bf u}_\perp, w)$ remains incompressible, $\nabla \cdot {\bf  u} = 0$, while small density variations are taken into account only in the buoyancy term. Such a model can be written as:
\begin{eqnarray} 
\partial_t {\bf u} + ({\bf u} \cdot \nabla) {\bf u} &=& - \nabla p - N\theta {\bf \hat z} + 
\nu \nabla^2 {\bf u}  + {\bf F} \label{momentum} \\
 \partial_t \theta  +  ({\bf u} \cdot \nabla) \theta &=& Nw + \kappa  \nabla^2 \theta  \ . \label{scalar}
\end{eqnarray} 
with $\theta$ being the temperature fluctuations evaluated relative to a mean temperature profile $\theta_0$, $N = [-g\partial_z {\bar \theta}/\theta_0]^{1/2}$ is the Brunt-V\"ais\"al\"a frequency. Such a simulation has been initialized with zero temperature fluctuations and random velocity modes applied at large-scale, in a Fourier shell centered at $k_0 = [2,3]$; the size of the computational box is associated with $k_{min}=1$, with $k_{min}=2\pi/L_0$ and the resolution is $512^3$ grid-points. A random forcing $F$ is imposed {on} the momentum equation at $k_F = 2\pi/L_f \in [2,3]$, continuously injecting kinetic energy into the system and allowing {it} to reach a turbulent stationary state. In the above equations, $\nu$ and $\kappa$  are the kinematic viscosity and the thermal diffusivity, respectively and we take the Prandtl number $\nu/\kappa=1$, with $\nu=10^{-3}$; finally, $p$ is the pressure.
 We adopt the following definitions for the dimensionless Reynolds and Froude numbers,
\begin{equation}
    \mathrm{Re}= U_{rms}L_{int}/\nu, \,\, \mathrm{Fr} = U_{rms}/[L_{int}N] \ , \label{EQ:RE}
\end{equation} 
with $U_{rms}$ and $L_{int}$  the characteristic ({root mean squared}) velocity and the integral scale of the fluid, respectively. 
These parameters, and in particular the buoyancy Reynolds $R_B=\mathrm{Re}\mathrm{Fr}^2$, measure the relative strength of buoyancy to dissipation and are commonly used to distinguish between wave-dominated and turbulence-dominated regimes~\citep{ivey_08}. 
The equations (\ref{momentum}--\ref{scalar}) are integrated numerically using the 
Geophysical High-Order Suite for Turbulence (GHOST), a pseudo-spectral code {that} employs a hybrid parallelization combining MPI, OPENMP{,} and CUDA \citep{mininni_11h, rosenberg_20}. It allows for a variety of physical solvers and supports non-cubic geometry \cite{sujovolsky_18}, and non-periodic boundary conditions~\citep{fontana_20}. 
In the following, we will make use of the statistical moments of a distribution function up to the {fourth-order}. The definition of the third- and forth-order moments, namely the skewness $Sk_w$ and kurtosis $K_w$ respectively, of the distribution of vertical velocity $w$ is:
\begin{equation}
Sk_w = \frac{\left<(w -  \bar{w})^3 \right>}{\left<(w - \bar{w})^2 \right>^{3/2}}, 
\label{EQ:SKw}
\end{equation}
\begin{equation}
K_w = \frac{\left<(w -  \bar{w})^4 \right>}{\left<(w - \bar{w})^2 \right>^2},
\label{EQ:Kw}
\end{equation}
where averages can be either computed over the {entire} simulation volume or {on sub-domains}, such as horizontal planes. It is worth recalling that $Sk=0$ and $K=3$ are the reference values for a Gaussian distribution, {whereas} values of $K_w$ {larger} than 3 {are indicative of} PDFs {with} fat (non-Gaussian) tails~\citep{Feraco2018}. 
As observed by~\citet{Feraco2018}, {the volume kurtosis }$K_w$ {shows large} non-Gaussian values in a narrow range of the Froude number, with {a} peak at $Fr \approx 0.076$, compatible with {actual geophysical flows}. {The study presented here focuses on a run characterized by this value of the Froude number,} in~\citep{Feraco2018,pouquet2019,Feraco2021,Marino2022}, {exhibiting indeed} {high} levels of {large-scale} intermittency (see Fig.2 of~\citet{Feraco2018}). {This simulation is well resolved, meaning the ratio between the largest wave number $k_{max}\approx 512/3$ and the Kolmogorov scale $k_\eta=(\nu^3/\epsilon_v)^{-1/4}$, $\epsilon_\nu$ being the kinetic dissipation rate, is $k_{max}/k_\eta\sim 1.8$.}

\begin{table}
\centering
\small
\begin{tabular}{rcccc}
\hline
 Id & Compression factor & No. Layers & No. params. & Latent space \\ \hline
CF2 & 2 & 7 & 321,156 & $64\times 64\times 128$ \\
CF4 & 4 & 8 & 339,892 & $32\times 32\times 256$\\
CF8 & 8 & 9 & 324,772 & $32\times 32\times 128$\\
CF16 & 16 & 11 & 349,804 & $16\times 16\times 256$\\
\hline\hline
\end{tabular}%
\caption{Description {of} the four CAEs architecture: the compression factor (CF), the number of layers referring to both the encoding and decoding part of the network. The number of parameters {accounts} for all the weights and biases {that} have to be optimized during the training phase, and the latent space is the reduced space after the encoding network is applied.}
\label{tab:tab_cae}
\end{table}

\section{\label{sec:cae_basics}Convolutional Autoencoder theory}

\begin{figure*}
    \centering
    \includegraphics[width=\linewidth]{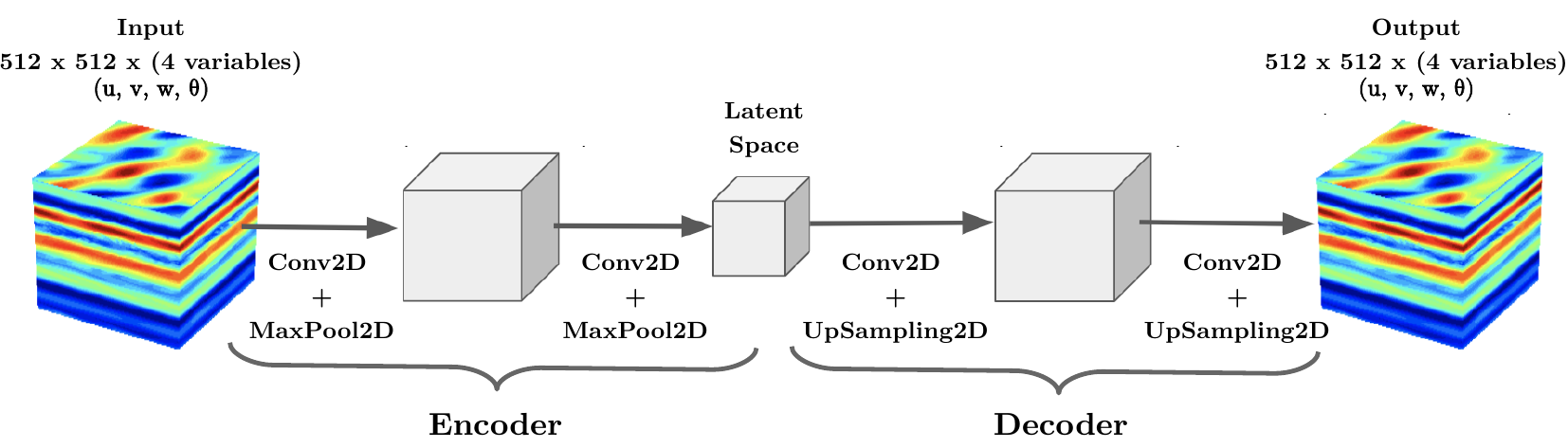}
    \caption{Schematic representation of a general Convolutional Autoencoder (CAE). The dimension of the \emph{latent space} and the number of encoding/decoding layers depend on the compression factor (CF). For each time step, the input is represented by horizontal slices ($512^2$) of the four physical variables produced by the simulations ($u$, $v$, $w$ and $\theta$) stacked together forming a volume of $512^2\times 4$. The number of filters and layers composing the encoder (and decoder) part also {depends} on the CF, all the {CAEs} implemented in this work are symmetric.}
    \label{fig:NetworkArchEx}
\end{figure*}

An autoencoder is an unsupervised {feed-forward} artificial neural network (ANN) that aims to reconstruct a given data set through a process involving data compression and recovery \citep{DeMers1993}. Indeed, its main objective is to create a reliable and reduced representation of the input data set, {which is} particularly helpful for either creating reduced {models} or simpler but still informative {representations} of high-dimensional data.
The network consists of an encoder part $P:\mathbb{R}^D\xrightarrow{} \mathbb{R}^d$ (with $d<D$), which compresses the input data into a smaller space representation (or latent space), and a decoder part $Q:\mathbb{R}^d\xrightarrow{} \mathbb{R}^D$, which reconstructs the encoded data back to the original input dimension. When the mapping kernel is linear, the autoencoder can be considered as a singular value decomposition analysis. An autoencoder that employs convolutional layers to perform the encoding and decoding operations on images, or more in general on 2D (or higher) data, is generally termed Convolutional AutoEncoder (CAE). During the convolution operation, the kernel swipes the data domain to extract features and learn spatial and/or temporal dependencies. The process is then carried out across multiple layers, obtaining representative features in a hierarchical sense~\citep{Masci2011}; CAEs with many hidden layers can be considered as deep neural networks (DNN). The initial layers learn a low-level representation or particular details about the data, while subsequent layers focus on larger and larger-scale information. In our work we developed four CAEs with different compression factors (CFs), defined as the ratio between the input and the latent dimensions, from 2 to 16; {this can be also seen as the ratio between the degrees of freedom of the initial system with respect to the reduced one, being generally proportional in numerical simulations to the number of grid points, therefore to the Reynolds number in the case of fluids.} An example of a general CAE architecture used in our implementations is shown in Fig.~\ref{fig:NetworkArchEx}. In Tab.~\ref{tab:tab_cae} we report the main characteristics of the four CAEs implemented here; starting from CF2, we obtained higher CFs by adding more hidden {layers} to the initial architecture but trying to keep the number of parameters fixed (as possible) in order to make a more fair comparison of the four networks performance. Indeed, the number of weights and biases, as highlighted by the values in Tab.~\ref{tab:tab_cae}, varies only by $\approx 10\%$. On the other hand, more hidden layers allow networks with higher CF to learn more features of the original fields during the compression step. The number of hidden layers is intended for the entire CAE architecture, comprising both the layers of the encoding $n_e$ and decoding $n_d$ part; they usually have {a number} of layers which follow the relation $n_d=n_e+1$, while the CAE is symmetric if the total number of layers is even.

\section{\label{sec:sec4}High-order statistics informed neural network: H-SiNN}
\begin{figure}[t]
    \centering
    \includegraphics[width=1\columnwidth]{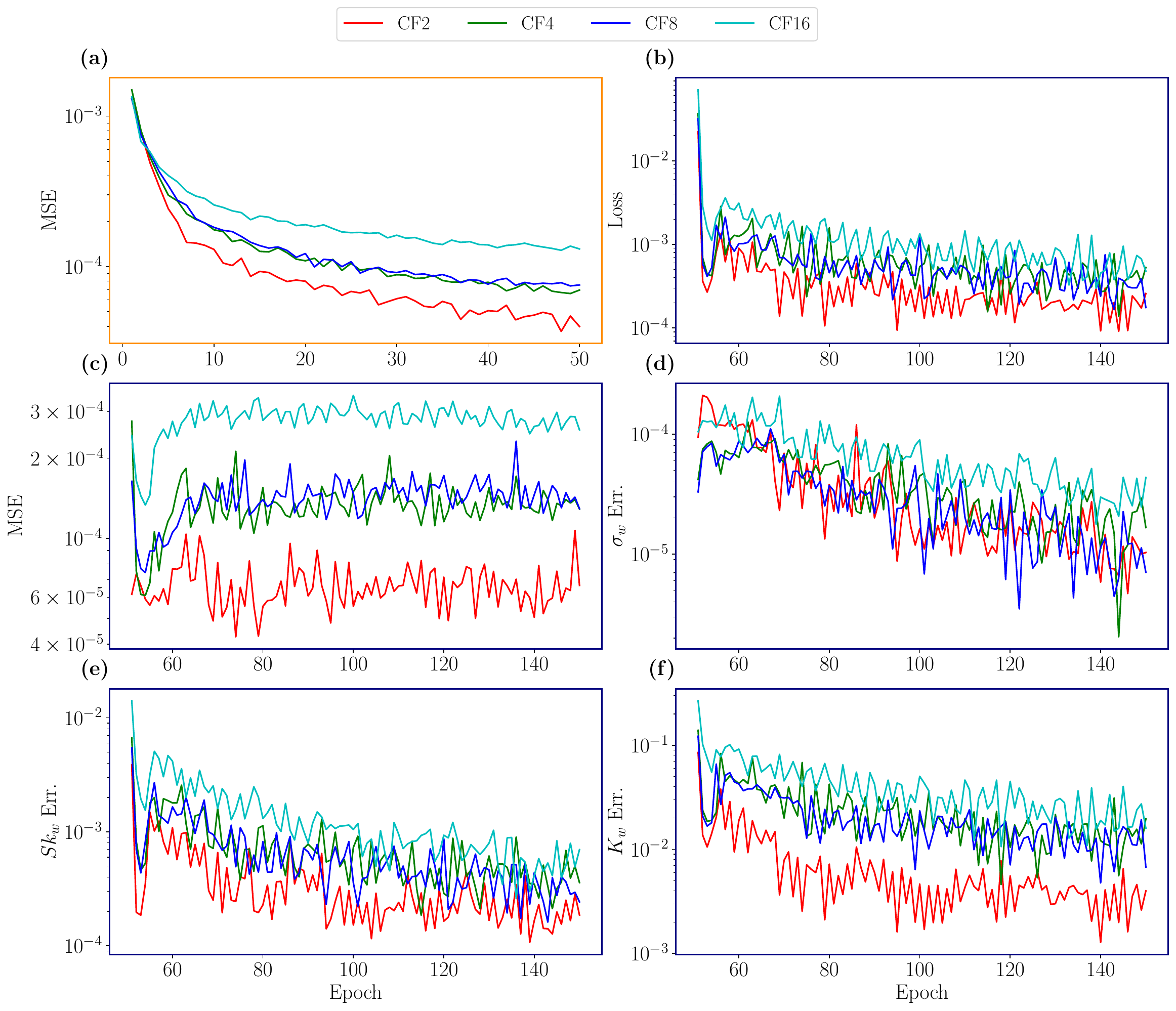}
    \caption{Panel (a) shows the mean squared error (MSE) during the first 50 epochs of the training stage where no other terms are added to the loss function. In panel (b) the total loss during 100 epochs of the second training stage, in which the loss function is composed by four terms which are represented in panels (c)--(f) for the various CAE with different compression factor (CF).}
    \label{fig:train_loss}
\end{figure}

The aim of this work is to devise a method that attains a compression of the data volume of DNS for stratified turbulence flows, {while} preserving their statistics (up to the fourth-order moment, kurtosis). Mathematically, this goal can be cast as a multi-objective optimization (MOO) problem and one of the greatest advantage of using a neural network to solve MOO problems is that, in principle, it is sufficient to include any term that one wishes to minimize in the cost function and, as long as such terms are differentiable, the neural network will find a solution that approximately minimizes the objective (possibly a local minimum) through the back-propagation procedure. This is known as \emph{scalarization method}, where the different terms {are} simply added together, each one weighted by its own scalar factor \citep{gunantara2018review}. Determining those weights is crucial, especially when employing ANNs, to make sure that the cost function is not dominated by one or a few terms only and that the solution actually minimizes each term. Intuitively one can think of the weights as normalization factors when different terms in the cost function have different {orders} of magnitude. 
Our implementation of H-SiNN involves the use of a statistics-aware loss function {in} a CAE, thus creating what we call statistics-informed convolutional autoencoder (SiCAE).
As mentioned above, we focus on up-to-fourth order moments of the vertical velocity distribution, since we already know that it presents the highest variability in our simulations. In particular, our aim is to preserve standard deviation, skewness, and kurtosis of the vertical component of the velocity field $w$, which we indicate with the symbols $\sigma_w$, $Sk_w$, and $K_w$, respectively. As a result, the cost function of the CAE computed between the original data set $y$ and its reconstruction $\tilde{y}$ is defined as:
\begin{equation}\label{eq:ch6_custom_loss}
\begin{aligned}
    \mathcal{L}(y,\tilde{y})&=\frac{1}{D}\Bigg(W_1\sum_{y}^{[u,v,w,\theta]}\sum_{i=1}^{M} \frac{(y_i - \tilde{y}_i)^2}{M} + W_2\sum_{i=1}^{M} \frac{(\sigma_{w_{i}} - \sigma_{\tilde{w}_{i}})^2}{M} \\
    & + W_3\sum_{i=1}^{M} \frac{(Sk_{w_{i}} - Sk_{\tilde{w}_{i}})^2}{M} + W_4\sum_{i=1}^{M} \frac{(K_{w_{i}} - K_{\tilde{w}_{i}})^2}{M}\Bigg)\\
    & = \frac{1}{D}\sum_{j=1}^4 W_j R_j
\end{aligned}
\end{equation}
with $D=3(W_1+W_2+W_3+W_4)$, being the total sum of the weights.
The second, third{, and fourth terms} in~\eqref{eq:ch6_custom_loss} are the mean square errors (MSEs) between the moments of the original and the reconstructed vertical velocity $W$ computed over horizontal planes ($x$,$y$), namely the standard deviation $\sigma_w$, the skewness $Sk_w$ and kurtosis $K_w$ respectively. We recall that for a normal distribution{,} the reference values are: $\sigma = 1$, $Sk=0$ and $K=3$. 
In addition, here, we propose a method for dynamically {weighing} the four terms in the cost function; each term is properly weighted with values $W_i$ varying at each epoch during the training phase. In order to obtain a loss function with terms having the same order of magnitude during the whole training procedure at a given epoch $m$, the weight of the $i$th term $W_i$ is proportional to the sum of the other three terms at the previous epoch $m-1$, e.g. $W^m_i=\sum_{j\neq i}W^{(m-1)}_j R^{(m-1)}_j$ (and normalized such that their sum is equal to 1). Even though the various terms of the statistics-informed loss function are well balanced, the training procedure with additional constraints either coming from the knowledge of the physical system, as it happens for physics-informed neural networks {(PINNs)}, or from statistical information can be difficult, slower{,} and sometimes very noisy.
{In order to mitigate gradient's pathologies when dealing with complex loss functions having several terms, a different approach consists in defining $W_i$ as proportional to the gradients of the $i$-th term with respect to the network parameters~\citep{Wang2020}. This technique was shown to be  particularly efficient in balancing the loss function, especially in PINNs~\citep{Wang2020,Leoni2023}. A claim was made in the referenced studies that using weights proportional to the loss itself leads to convergence issues, though it does not seem to be the case in our application, where the various loss terms are well balanced and the global loss trend decays properly (see Fig.~\ref{fig:train_loss}). That being said, a comparison between the two approaches would be interesting as a future study as it may lead to further improve the novel CAE we propose here.\\
Finally, we have heuristically verified that the strategy of training an H-SiNN in two stages is computationally faster and more robust.} In the first stage we treat the autoencoder as a classical CAE, with only the first term in Eq. \ref{eq:ch6_custom_loss}. After we train for 50 epochs, we add the remaining terms in the cost function, i.e. starting from a network that achieves a low reconstruction error, in terms of mean error.
During the second stage{,} we train the CAEs for 100 more epochs with the addition of statistical constraints on the first four moments of the vertical velocity PDF; this is shown in panels (b)-(f) of Fig.~\ref{fig:train_loss}.
{A similar strategy has been successfully implemented for a variety applications, most of them in the context of PINNs~\citep{Camporeale2022,Leoni2023}.}
The behavior of the different terms of the loss function during one hundred epochs of training is shown in Fig.~\ref{fig:weights_his}. We can appreciate from the four panels how there are essentially three weights ($W_{MSE}$, $W_{\sigma_{w}}$ and $W_{Sk_{w}}$) which are nearly equal for the entire training phase, whereas the coefficient weighting the kurtosis term is significantly smaller indicating a larger error on the forth-order moment, as expected.

\subsection{Plane-by-plane approach}
We mentioned that the main objective is to obtain a tool {that} is able to reproduce most of the features of the velocity $\mathbf{u}$ and temperature $\theta$ fields of a fluid presenting high anisotropy (stratification) and non-homogeneity, due to the presence of large-scale intermittent structures. For these last reasons, we further implemented the standard CAE algorithm to make it more suitable and better performing in our particular application. A key ingredient of stratified flows is anisotropy, introduced by gravity, which {tends} to {suppress} vertical motions. Therefore, even though we deal with fully three-dimensional simulations the original cubic volume is split in horizontal planes along the vertical direction ($z$) before being used as input for the CAEs. This means that, for each time and height value $z$, planes for every velocity component ($u$, $v$ and $w$) and for $\theta$ are stacked together creating three-dimensional input data with dimensions $512\times512\times4$, as indicated in Fig.~\ref{fig:NetworkArchEx}. 
Before being divided, the data are normalized $\hat{x}=(x-\mu)/\sigma$ using the average $\mu$ and standard deviation $\sigma$ computed on each simulation cube. In this approach, we believe that passing the information from the three components of the velocity, together with the temperature, is essential for the CAE to better reconstruct the flow dynamics at a given time $t$ and altitude $z$ since their variation is strictly correlated within the primitive equations. {This approach is limited in the possibility to add important information about the velocity field, like the compressibility condition} $\boldsymbol{\nabla} \cdot \mathbf{u}=0$, since {operating} by-plane {poses} constraints on the vertical derivative of the velocity $\partial_z w$. This leads to values for the maximum and average divergence of $\mathbf{u}$ reported in Tab.~\ref{tab:divergence}, where we can see that even if on average the condition is well satisfied in the reconstructed fields, locally this is not true while the condition is satisfied also point-wise for the original field.

\section{\label{sec:results} Results}
We analyze in {detail} the output of a single simulation run for more than $45\tau_{NL}$ corresponding to 100 time steps. Since in our approach we divide each simulation cube in planes along the vertical direction which are then treated as independent during training, the data set we use for training and testing our CAEs comprises {$N_s=51,200$} samples, each with dimension $512^2\times 4$ ($512^2$ grid-points for four variables $u$, $v$, $w$ and $\theta$). We divide the data set in training and test set on a temporal basis, taking the first 70\% snapshots for the training phase and the remaining 30\% for testing the performance.
\\
The {above strategy} has been used to train four different statistics-preserving CAE, with compression factors (CF) of 2, 4, 8, and 16. Figure \ref{fig:train_loss} shows the history of the different terms in the loss function, for {an} increasing number of epochs. Red, green, blue, and cyan lines are respectively for CF 2, 4, 8, and 16. The first stage is made of 50 epochs, and in the second stage{,} the networks are trained for an additional 100 epochs. 
The interesting feature, valid for all CFs, is that after the first stage (panel a) the MSE does not further {decrease} (panel c), while all other moments decrease by one order of magnitude or so (panels d-f). This shows that the second stage of training achieves the goal of minimizing the error of standard deviation, skewness, and kurtosis, without significantly {degrading} the point-wise accuracy of the reconstruction.

\begin{figure}
    \centering
    \includegraphics[width=1\columnwidth]{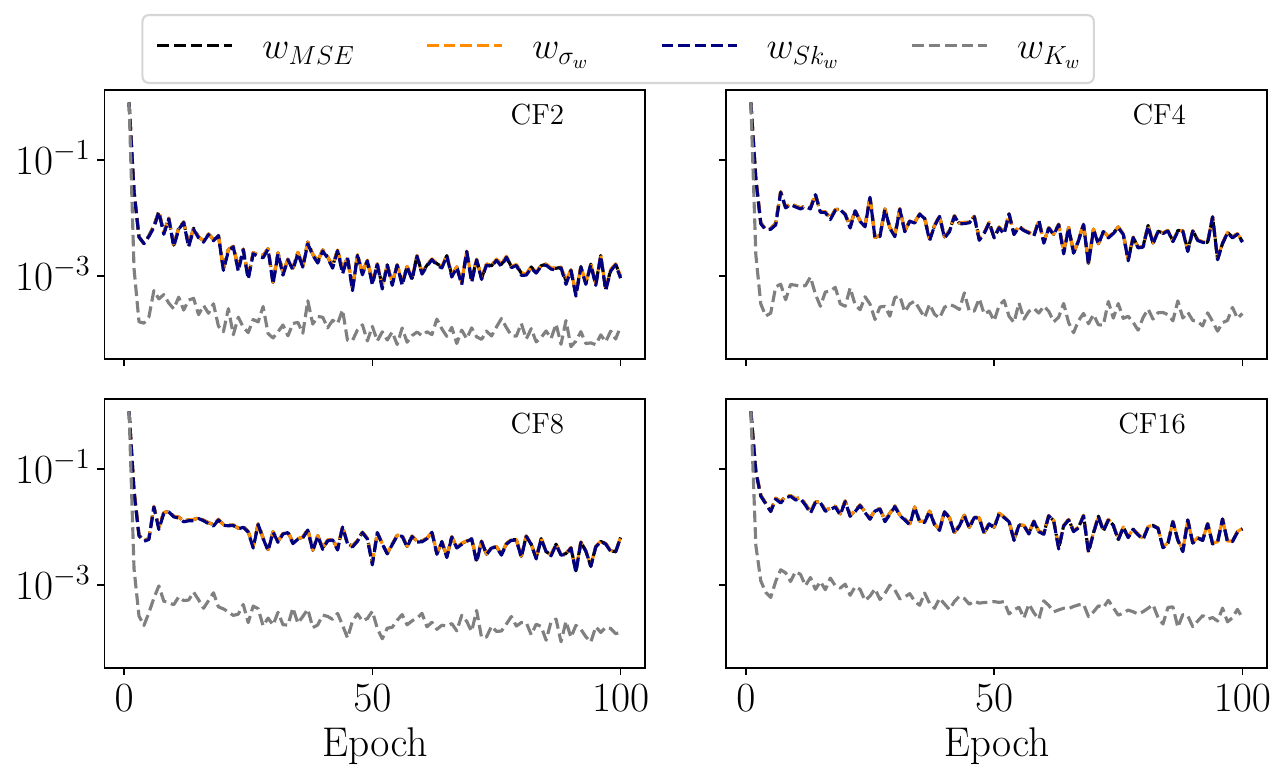}
    \caption{History of the weights multiplying the different terms of the loss functions: the mean squared error (MSE, black), the standard deviation ($\sigma_w$, orange), the skewness ($Sk_w$, blue) and the kurtosis ($K_w$, gray).}
    \label{fig:weights_his}
\end{figure}

\begin{table}[b]
\centering
\begin{tabular}{c|c|cccc}
 & Original & CF2 & CF4 & CF8 & CF16 \\ \hline \hline
$|\langle \boldsymbol{\nabla} \cdot \mathbf{u}\rangle| \;\;\; [\times 10^{-11}]$ & $0.42$ & $1.83$ & $3.26$ & $2.37$ & $2.06$ \\
$max \left\{|\boldsymbol{\nabla} \cdot \mathbf{u}|\right\}$ & $5.9\cdot 10^{-4}$ & $538$ & $410$ & $573$ & $530$ \\
\hline \hline
\end{tabular}%
\caption{Average and maximum value of the divergence of the velocity field after the reconstruction with the different CAE.}
\label{tab:divergence}
\end{table}

\subsection{Global statistical properties of reconstructed fields}
The trend of the statistical moments error {appears} to be inversely proportional to the moment {order}, meaning that the improvement on the reconstructed standard deviation (2nd-order moment) is {more significant} than what is achieved for the kurtosis (4th-order moment) within the same amount of epochs. This is also pointed out in Fig.~\ref{fig:weights_his} where, for the four different SiCAE, the trend of the various weights balancing the loss function terms is reported as a function of the epoch; let us recall that each weight $W_i$ is inversely proportional to the sum of the three {contributions} given from the other moments, taken at the previous epoch $~1/\sum_{(i\neq j)}W_j$.
In addition, by looking at the behavior of the loss weights (Fig.~\ref{fig:weights_his}), we observe that the error on the kurtosis is approximately between one and two orders of magnitude {larger} than the others. {This significant} difference may be due to the high variability of $K_w$ in this simulation\cite{Marino2022}. {Instead,} standard deviation and skewness are expected to vary less and to be closer to reference values of a Gaussian distribution.
{As it is reasonable,} a trend with the compression factor is observed, and as expected the higher CF the larger the reconstruction error both on the mean field and the statistical moments; this is not related to the newly introduced loss function since already during the first training stage (panel (a)) this trend can be clearly observed. 
The details of the results obtained on the test set for the various terms involved in the loss function during the two training stages are summarized in Tab.~\ref{tab:tab_results}. We can see that including other terms into the loss function results in an average reconstruction error 2-3 times greater than what {is} obtained after the first training stage, even though it remains on the order of MSE$\sim 10^{-4}$, consistent with other results found in literature~\citep{Glaws2020,Mohammadreza2022}. In addition, when constraints on the statistical moments are added to the model, we observe that the reconstruction of high-order moments improves up to 10 times for any compression factor.

\begin{table}
\centering
\resizebox{\columnwidth}{!}{%
\small
\begin{tabular}{lcccccc}
 &  CAE & MSE & $\sigma_w$ err. & $Sk_w$ err. & $K_w$ err. & Loss\\
 &  & $[\times 10^{-4}]$ & $[\times 10^{-5}]$ & $[\times 10^{-4}]$ &$[\times 10^{-2}]$ & $[\times 10^{-4}]$\\ \hline
1st stage\\``standard loss'' &  &  &  &  &  &  \\ \hline
 & CF2 & 0.39 & {\textit{0.13}} & {\textit{10.9}} & {\textit{18.4}} & 0.39 \\
 & CF4 & 0.55 & {\textit{0.45}} & {\textit{13.2}} & {\textit{22.4}} & 0.55 \\
 & CF8 & 0.54 & {\textit{0.82}} & {\textit{19.1}} & {\textit{19.0}} & 0.54 \\
 & CF16 & 1.1 & {\textit{5.3}} & {\textit{44.9}} & {\textit{41.7}} & 1.1 \\ \hline
2nd stage\\``statistical-informed loss'' &  &  &  &  &  &  \\ \hline
 & CF2 & 0.67 & 1.03 & 1.9 & 0.04 & 0.44 \\
 & CF4 & 1.0 & 1.0 & 3.2 & 1.4 & 4.0 \\
 & CF8 & 1.3 & 0.074 & 3.6 & 1.7 & 2.3 \\
 & CF16 & 2.3 & 0.32 & 12.0 & 5.9 & 10.0 \\ \hline\hline
\end{tabular}%
}
\caption{Results of the four CAE on the test set for both the first (top) and second (bottom) training phase. {Values in italic represent only check quantities monitored during the first training phase, though they are not included in the loss function at this stage.}}
\label{tab:tab_results}
\end{table}

\begin{figure}
    \centering
    \includegraphics[width=1\columnwidth]{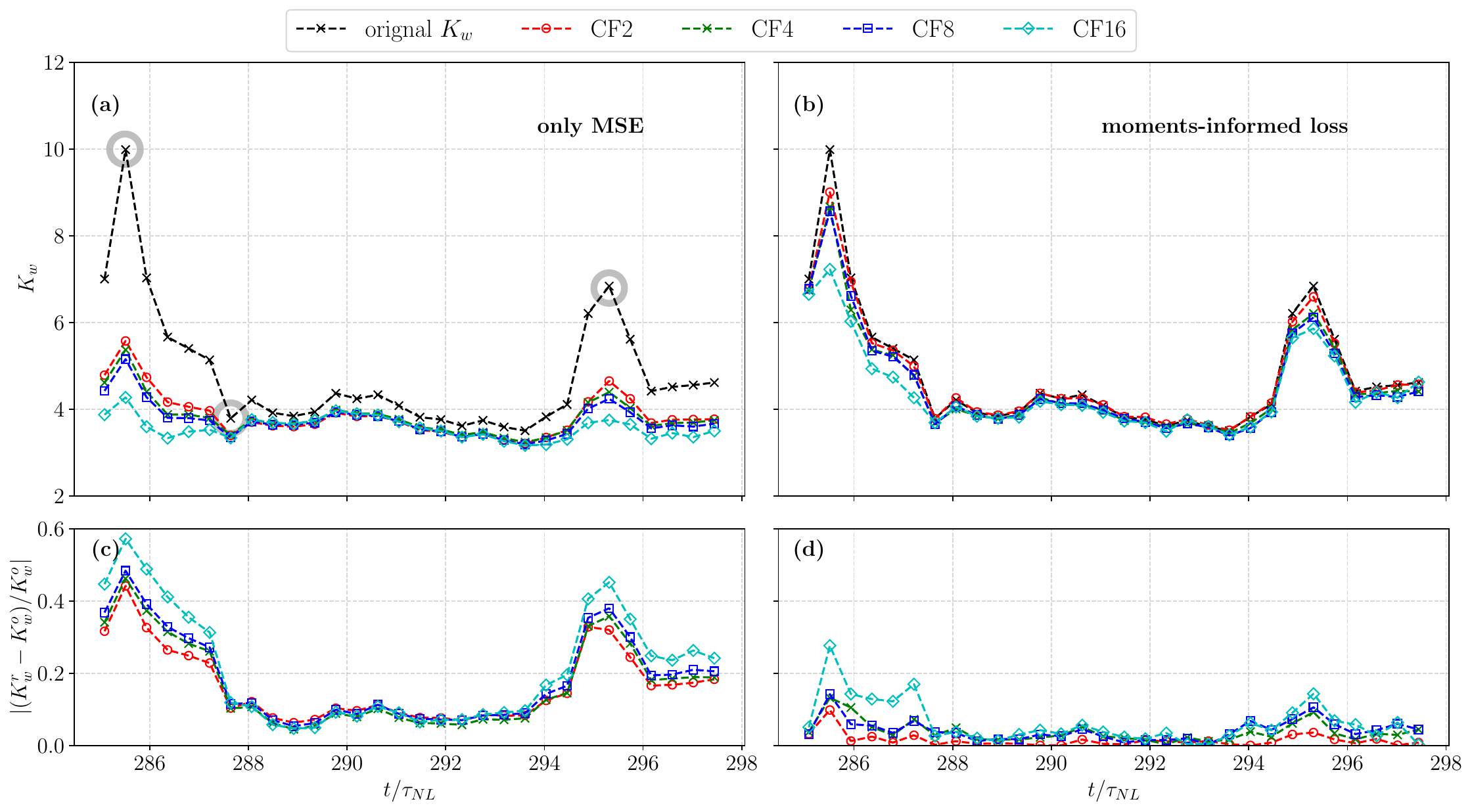}
    \caption{Behavior of the kurtosis of the vertical velocity $K_w$ over time for the test set. Panels (a) and (c) show $K_w$ obtained for the different compression factors (CFs) in different colors (top) and the mean absolute relative error on the reconstructed kurtosis (bottom) during the first training stage when the loss function is composed by the MSE only. The same results are shown in panels (b) and (d) for the second training stage where the modified loss is adopted. Gray circles in panel (a) represent three times analyzed in the following figures with very different values of kurtosis $K_w$ of the vertical velocity.}
    \label{fig:Kw_vs_t}
\end{figure}

\begin{figure}
    \centering
    \includegraphics[width=0.7\columnwidth]{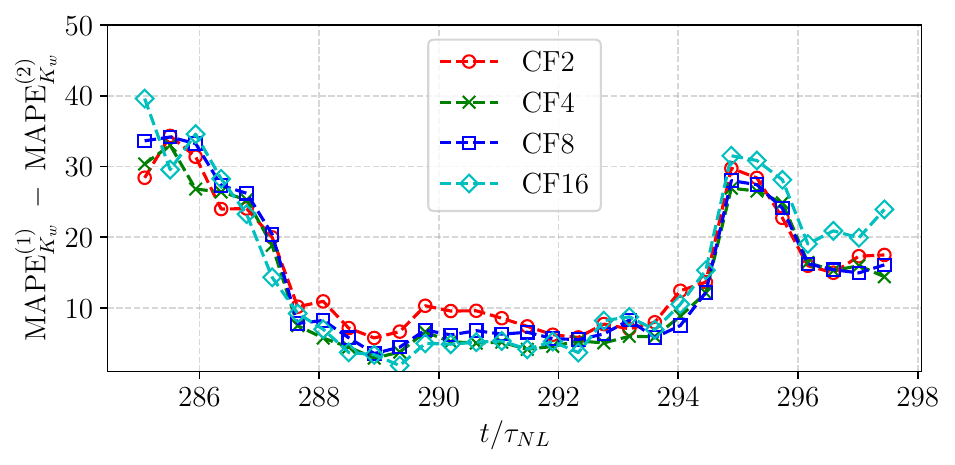}
    \caption{For each CAE, the behavior over time of the difference between the mean absolute percentage errors (MAPE) on the kurtosis $K_w$ of the vertical velocity between the first $\mathrm{MAPE}^{(1)}_{K_w}$ and second training stage $\mathrm{MAPE}^{(2)}_{K_w}$.
    The MAPE is defined as $\mathrm{MAPE}^{(i)}_{K_w}=|(K_w^{r,(i)}-K_w^{o})/K_w^{o}|\times 100$ with $K_w^{o}$ and $K_w^{r,(i)}$ being the kurtosis of the original and reconstructed fields, respectively.}
    \label{fig:Kw_error_ratio}
\end{figure}

\begin{figure}
    \centering
    \includegraphics[width=1\columnwidth]{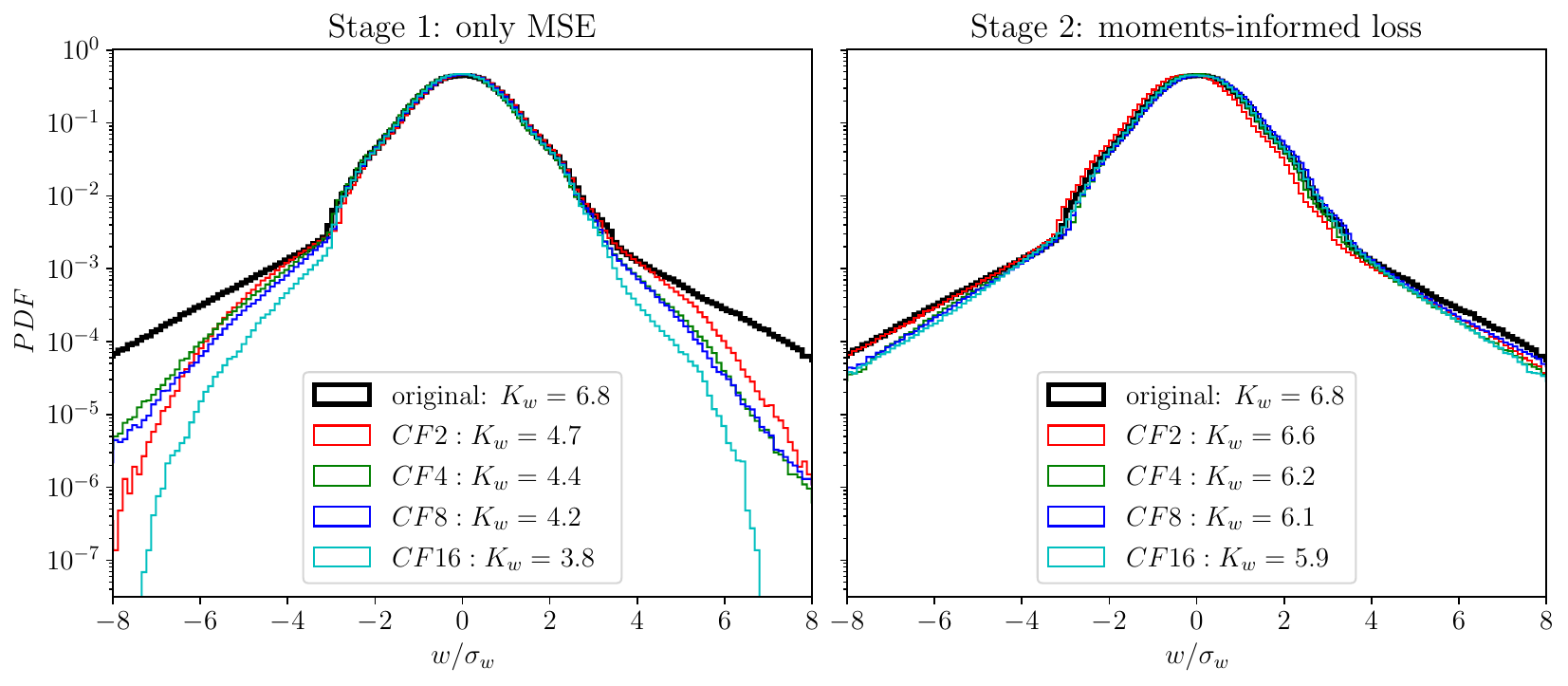}
    \caption{PDF of the vertical velocity $P(w)$ computed after the first training stage (left panel) and at the end of the entire training stage (right panel), taken at {$t/\tau_{NL}\approx 295$}. {One can appreciate the PDF core being in excellent agreement during stage one, while the fat tails of the distribution are only recovered after introducing the custom terms to the loss function in the second training stage (right panel).}}
    \label{fig:PDF_Vz_0694}
\end{figure}

\begin{figure*}
    \includegraphics[width=1\textwidth]{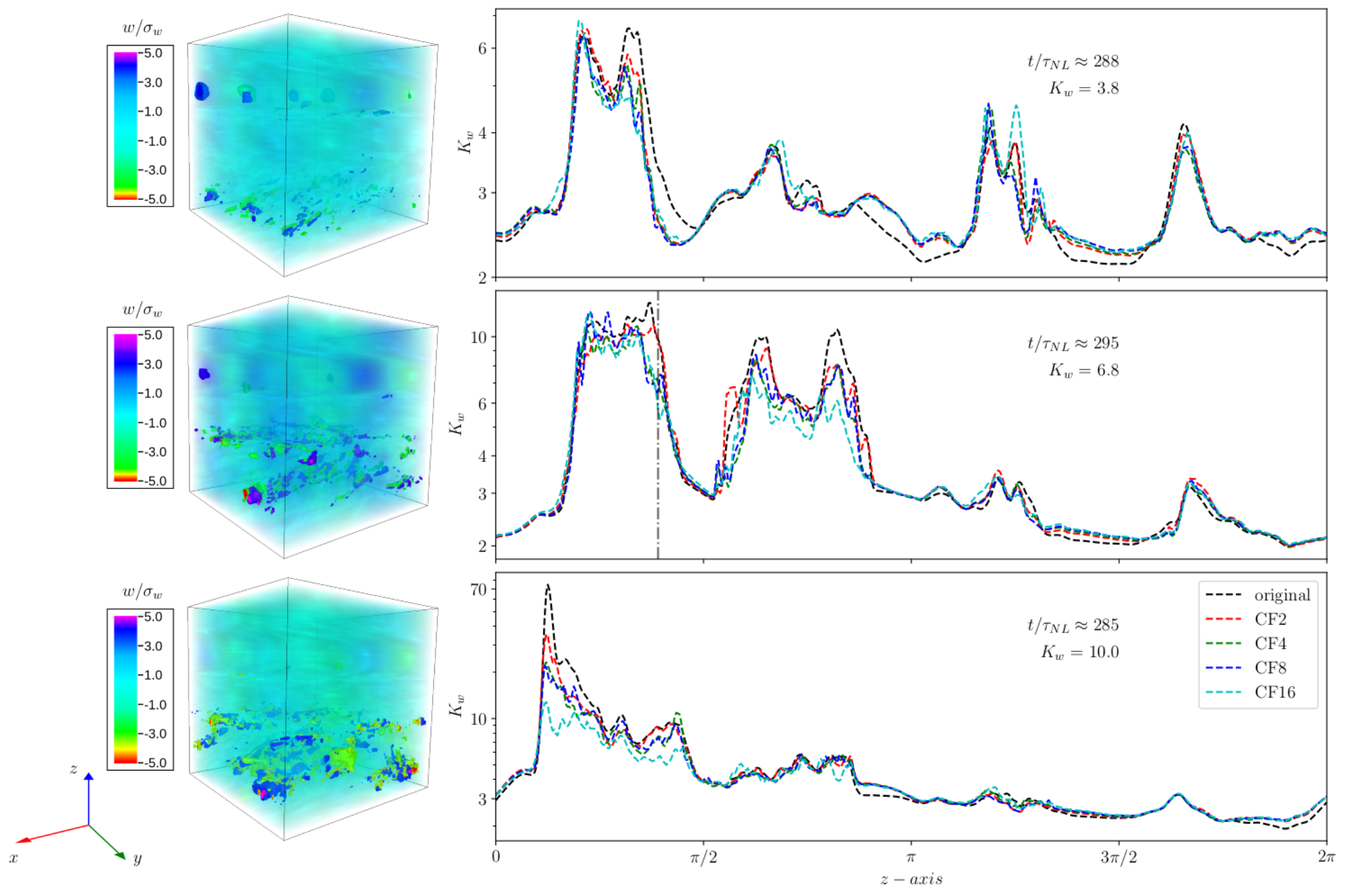}
    \caption{Left column: 3D renderings of the vertical velocity $w/\sigma_w$, where the most extreme structures have been highlighted in solid colors. Right column: Vertical profile of the vertical velocity kurtosis $K_w(z)$ of the original field (black dashed) and of the fields reconstructed with the SiCAE (colored, see legend). The vertical gray dash-dotted line (mid panel) {represents} the time and height taken for Fig.~\ref{fig:2D_rendering_allFields}. Let us notice that the kurtosis is in logarithmic scale and the three panels show a different y-range for a clearer visualization.}
    \label{fig:Kw_per_plane_all}
\end{figure*}

\begin{figure*}
    \includegraphics[width=1\textwidth]{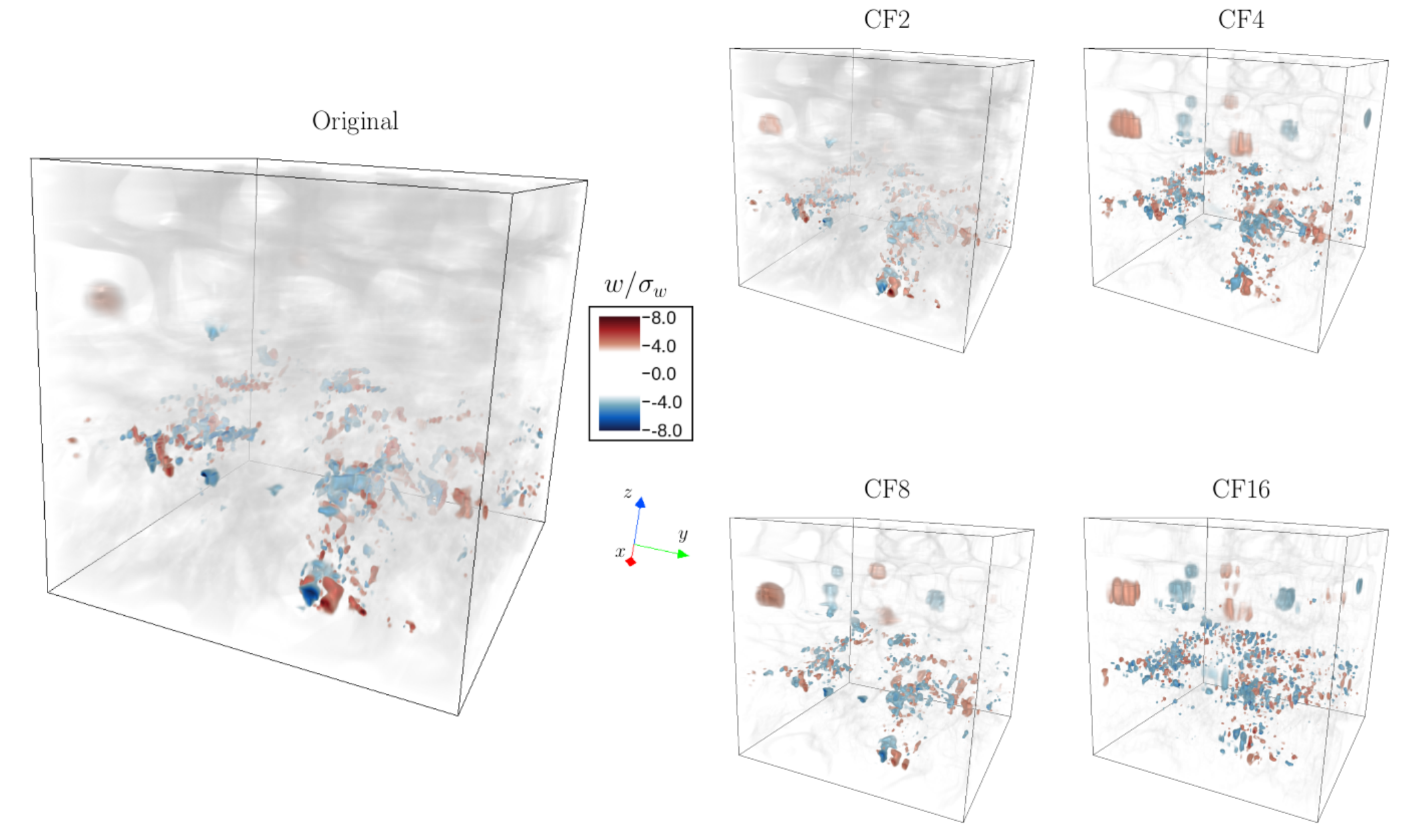}
    \caption{Three-dimensional visualizations, taken at {$t/\tau_{NL}\approx 295$}, of the extreme events in the vertical velocity $w$ from the original data (big cube), alongside the field reconstructed by the four SiCAE (small cubes).
    The vertical drafts ($|w/\sigma_w|>4$) are highlighted as solid colors (red/blue for positive/negative), whereas the other regions are represented as {gray-shaded} areas}
    \label{fig:Vz_3D}
\end{figure*}

\begin{figure*}
    \includegraphics[width=1\textwidth]{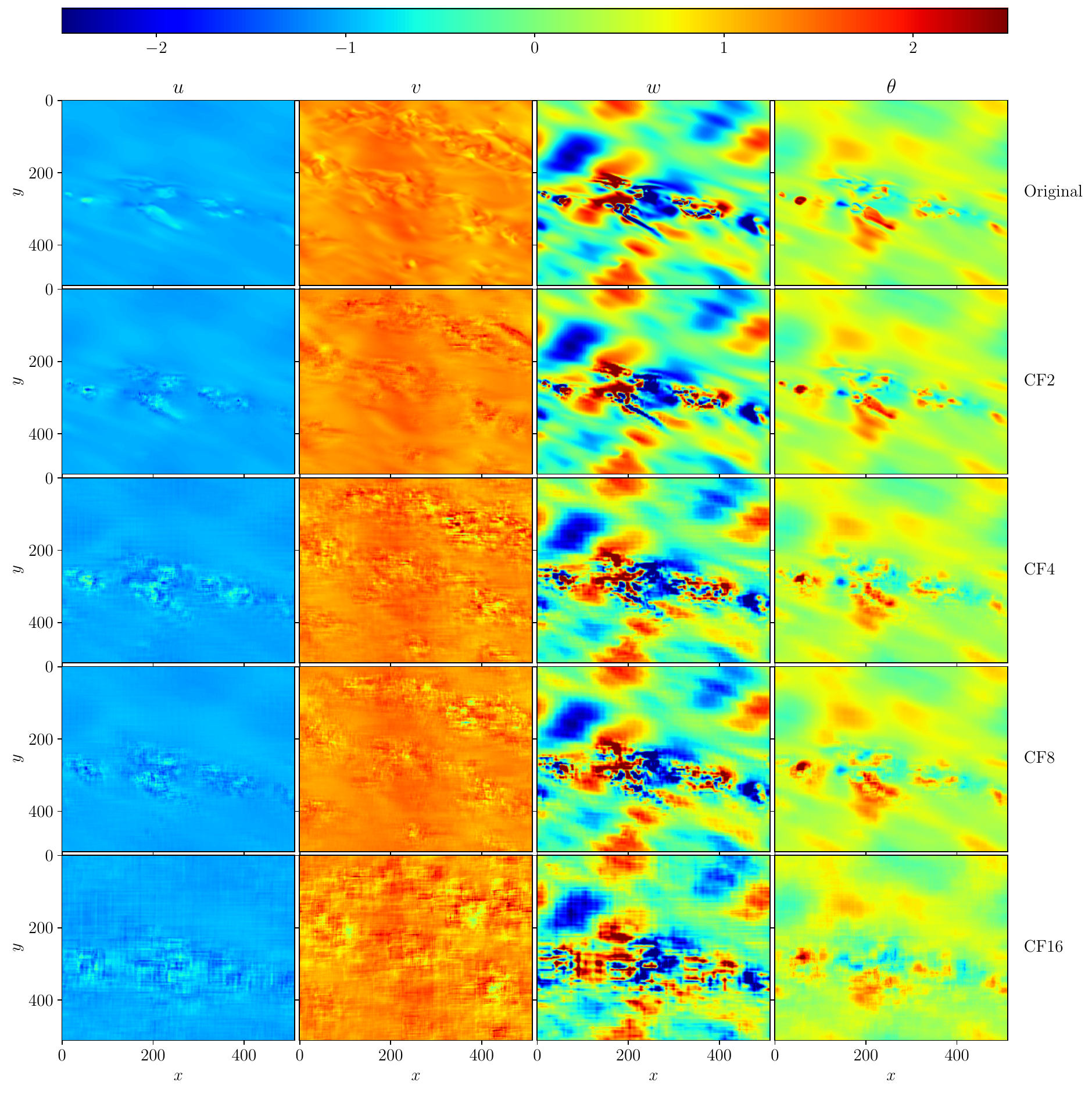}
    \caption{Horizontal slices of the four physical fields, from left to right: horizontal component of the velocity $u$ and $v$ (left and mid-left), the vertical component $w$ (mid-right) and the potential temperature $\theta$  (right) taken at $t/\tau\approx 295$ with a kurtosis $K_w(z^*)=10.0$. From the second line to {the} bottom the reconstructed field obtained with different CAE having increasing compression factor (CF). 
    }
    \label{fig:2D_rendering_allFields}
\end{figure*}

We selected an interval of the numerical simulation where isolated extreme bursts occur, producing evident peaks of the kurtosis $K_w$ (see Fig.~\ref{fig:Kw_vs_t}, panel (a)). In Fig.~\ref{fig:Kw_vs_t} the vertical velocity kurtosis computed on the original data (black dashed curve) is shown as a function of the turnover time for the entire test set, and compared with the kurtosis obtained from the CAEs reconstructions (different colors for different CF values). Panels (a) and (b) show a comparison obtained respectively after the first and second training phase; the relative error between the original $K_w^a$ and recovered $K_w^r$ kurtosis is shown in panels (c, first phase) and (d, second phase). 
This figure clearly highlights how the second stage of {the} training (panels (b) and (d)) significantly improves the reconstruction of the fields up to the {fourth-order} moment of the vertical velocity. 
All the CAEs show a significant reduction of the relative error, with {marked} improvements at {the} times {at which} the volume $K_w$ exhibits large non-Gaussian values {($t\approx 285\tau_{NL}$ and $t\approx 295\tau_{NL}$)}. In addition, with the application of the modified loss we are able to obtain percentage errors on the kurtosis smaller than 10\% on average. However, the improvement in the reconstruction of the {fourth-order} moment is obtained also for values close to the Gaussian reference ($K_w\approx 3$). Indeed, it is evident that our approach can be successfully extended to other situations where the PDFs of the dynamical fields are far from Gaussian-like shape, as is the case for the concentration of a passive scalar~\citep{Pumir1991} or for the velocity in the boundary layer~\citep{Mahrt1987,Brun2017}. 
The statistical moments up to the {fourth-order} embedded into the loss function guarantee an optimal reconstruction of the statistical properties of the physical fields without any prior knowledge about the shape or variability of their distribution function. The improvement we obtain with the custom loss is quantified in Fig.~\ref{fig:Kw_error_ratio} where we show the difference between the mean absolute percentage error (MAPE) computed on $K_w$ after the first and second training stage, as a function of time (once again, note that the training is performed by treating each plane and each time independently, i.e. no correlation along the $z$ direction or in time is taken into account). The improvement reaches up to $\approx 40\%$ at times with high values of $K_w$ but in general{,} an {increase} of nearly $10\%$ is observed at all times. In Fig.~\ref{fig:PDF_Vz_0694} we show the PDFs computed on the entire volume at time {$t\approx 295\tau_{NL}$} (third gray circle in Fig.~\ref{fig:Kw_vs_t}), when $K_w\simeq 6.8$, using both the original data (black) and data obtained from the four CAEs (colored curves). We notice how after the first training stage (left panel) the PDF core is already reliably recovered by the decoder, even though the tails significantly differ. In particular, the difference between original and reconstructed {statistics} seems {to increase} for higher $|w|$. The PDF in the right panel of Fig.~\ref{fig:PDF_Vz_0694}, obtained after the second training stage, completely resembles the one computed from the original data, confirming that enforcing the statistical moments up to the forth-order is sufficient to enforce compatible PDFs.
In Fig.~\ref{fig:Kw_per_plane_all} we show an example of large-scale intermittent structures observed at three different times, indicated by the gray circles in Fig.~\ref{fig:Kw_vs_t}. The three-dimensional renderings represent the vertical velocity field $w$ for low {(top, $t\simeq 288\tau_{NL}$), medium (center, $t\simeq 295\tau_{NL}$) and high (bottom, $t\simeq 285\tau_{NL}$)} values of the kurtosis $K_w$, where values $|w/\sigma_w|>4$ are highlighted in solid color {while} smaller values are depicted as a transparent blue. Alongside each rendering the vertical profiles of the kurtosis $K_w$ computed plane-by-plane {on} the original data (black), as well as {on the} reconstructed {field} (colored for the different CFs) {are reported}. Also for the top panel, when the kurtosis is small $K_w \simeq 3.8$ there are planes reaching values of $K_w\approx 6$, thus indicating a strong spatial variability even when the global kurtosis is close to the Gaussian reference value. These effects are enhanced when more vertical drafts develop within the flow (center and bottom) with kurtoses more than one order greater than the Gaussian reference values, as shown {by the} logarithmic scale in Fig.~\ref{fig:Kw_per_plane_all}.

\subsection{CAE reconstruction of intermittent structures}
We observed that the statistical properties and in particular the kurtosis of the vertical velocity {of the flow under study} are well recovered by the proposed implementation of statistics-informed CAEs. {As already mentioned,} the presence of large-scale extreme events {produces intermittent} patches of enhanced turbulence and regions populated by transient coherent structures, {making the reconstruction of the fields and the associated non-Gaussian statistics challenging}. 
In this section{,} we analyze how the introduction of global statistical terms to the loss function {does improve the point-wise} reconstruction of the {full} velocity ($\mathbf{u}$) {and} potential temperature ($\theta$) {field over the entire domain, also by means of visualizations}. In Fig.~\ref{fig:Vz_3D} we report three-dimensional {renderings} of the vertical velocity $w$ obtained with the four CAEs, as well as the original field for comparison. Data are taken at time {$t\simeq 295 \tau_{NL}$}, {characterized by} relatively high kurtosis, $K_w\approx 6.8$. Extreme vertical drafts {are localized in the regions of the flow characterized by very large values of the original normalized vertical velocity ($|w/\sigma_w|$), in Fig.~\ref{fig:Vz_3D} these correspond to the domain points with $|w/\sigma_w|>4$}. The same kind of structures {emerge} in the renderings {of the reconstructed field}, up to the {highest compression level} (i.e., CF16); {though for} $CF=16$ the shape and location of {the regions characterized by vertical} drafts {appears blurred in the visualization compare to} the original field, {good spatial correlation can still be appreciated}. The gray transparent shading represents velocity values below the threshold of $4\sigma_w$, and therefore the majority of the volume. The reconstruction made by the implemented CAEs involves also the other components of the velocity field ($u$ and $v$), as well as the (potential) temperature fluctuations $\theta$. In order to have a general overview of how the neural networks recover all the physical quantities interested in the analyzed DNSs we represent them with several panels in Fig.~\ref{fig:2D_rendering_allFields}. This figure shows {a} horizontal cut ($x$,$y$) of the simulation domain taken at the same time {as} the previous figure (Fig.~\ref{fig:Vz_3D} at the height $z^*$ indicated with a dash-dotted line in Fig.~\ref{fig:Kw_per_plane_all} (middle panel). The columns of Fig.~\ref{fig:2D_rendering_allFields} refer to the three components of the velocity field $\mathbf{u}=(u,v,w)$ and the temperature $\theta$. All the images are represented by the same color bar which is not shown since the main objective of this figure is the comparison between the first row, the original data, and the others, being the reconstructed physical fields for increasing values of the compression factor. Since we are looking at the domain from {the} top (gravity is a vector entering the page), it is correct to have horizontal components strongly dominated by a nearly constant positive (red) for $v$ and negative (blue) for $u$ mean wind (see color bar); this is indeed the effect of Vertical Sheared Horizontal Winds (VSHWs) which are ubiquitous in stratified flows. Nevertheless, the horizontal components of the velocity show small-scale perturbations where extreme vertical drafts develop, as already seen throughout this manuscript. The extreme events developed in this snapshot are clearly visible in the vertical velocity $w$ and partially from the temperature renderings (third and {fourth} columns in Fig.~\ref{fig:2D_rendering_allFields}), and the same detail is captured also by the field reconstructions. Indeed, as already noticed for the three-dimensional visualization, the reconstruction is very reliable up to $\mathrm{CF}=16$ where a significant checkerboard effect starts developing everywhere in the domain. This is probably due to the combined effect of the high compression factor and of the statistical-informed loss function presenting large-scale statistical constraints. In fact, by looking at the reconstructed fields after the first stage of training (not shown here), {one} can observe the same artifact at high compression, {though} slightly reduced {due to} the absence of other terms {in} the loss function.

\section{Conclusion}
The {outcome of this work demonstrates how} the capability of machine learning to reproduce {dynamical} fields in fluid mechanics can be enhanced by incorporating physical and/or statistical knowledge {of} the system under study. In particular, we proposed a novel implementation of convolutional autoencoder (CAE) {that includes} a modified loss function able to preserve the statistical moments of the vertical velocity field up to the {fourth} order. We focused on the vertical component of the velocity field produced by DNS of {the Boussinesq equations with stable density stratification, since in a certain range of the parameters of geophysical interest - of the Froude number in particular - these stratified flows develop} strong velocity drafts along the $z$ direction, as observed in the atmosphere and oceans. {The existence of this phenomenology resulting from the interplay of turbulent motions and internal gravity waves, represents} a challenge {when it comes to find} an informative low-order representation of the physical fields. The comparison between a standard implementation of CAE and the SiCAE shows how the introduction of additional terms in the cost function {to enforce} high-order moments of the PDF, significantly {improves} the reconstruction of peculiar features of stratified geophysical flows, while maintaining an acceptable level of the error in the reconstruction of the mean fields.
The diagnostics implemented {here} confirm that the novel algorithm we proposed is able to reconstruct with good accuracy physical fields characterized by highly variable PDFs. In the framework considered, the emergence of vertical drafts {affects} the local dynamics of stratified turbulent fluids, {which end up being both non-stationary and non-homogeneous, making it more difficult to model their statistical description}. 
{The} results {we obtained} can be potentially generalized to other {physical fields and systems, and used} to address for instance the reconstruction of intermittent passive scalar fields~\citep{Orsi2021}, in order to investigate their diffusion properties, or the velocity field obtained from observations of the planetary boundary layer~\citep{Mahrt1987,Brun2017} and the upper atmosphere \citep{Vierinen2019,Chau2021}. {Highly} variable PDFs have also been {obtainted from the analysis of dynamical fields measured in the oceans} by means of GOMs~\citep{Pearson2018}, and using data from observations \citep{DAsaro2007}, further widening the application of the analysis proposed here, which can be very helpful in managing large volumes of geophysical data from high-resolution numerical simulations and state-of-art observational campaigns. The addition of other terms to the loss function {requires} to have longer training phases and larger dataset, in order to converge to a well-trained model. Indeed, very often a compromise {is} made between having detailed reconstructions of the mean {field} dynamics and reproducing peculiar structures of the turbulent flows, which is precisely where the algorithm {we} presented {improves} the classical methodological approach based on CAE. 

\begin{acknowledgments}
R.M. and R.F. acknowledge support from the project ``EVENTFUL'' (ANR-20-CE30-0011), funded by the French ``Agence Nationale de la Recherche'' - ANR through the program AAPG-2020. 
E.C. was partially supported by NASA grants 80NSSC20K1580 ``Ensemble Learning for Accurate and Reliable Uncertainty Quantification" and 80NSSC20K1275 ``Global Evolution and Local Dynamics of the Kinetic Solar Wind". {The simulations were ran on HPC facilities at the \'Ecole Centrale de Lyon (PMCS2I), in \'Ecully (France)}.
\end{acknowledgments}

\section*{Data Availability Statement}
The data that support the findings of this study are available from the corresponding author upon reasonable request.

\bibliography{aipsamp}

\end{document}